\documentclass[conference]{IEEEtran}

\ifCLASSINFOpdf
 
\else
 
\fi

\usepackage{array}
\usepackage{amsthm,amsmath,amssymb}
\usepackage{mathrsfs}
\usepackage{multirow}
\usepackage{siunitx}

\usepackage{threeparttable}
\usepackage{booktabs}
\usepackage{enumerate}
\usepackage{cite}
\usepackage{graphicx}
\usepackage[table]{xcolor} 
\usepackage{colortbl}
\usepackage{notation}
\usepackage{soul}
\usepackage{subcaption}
\usepackage{caption}
\begin{document}
\newtheorem{Def}{Definition}
\newtheorem{Pro}{Proposition}
\newtheorem{Lem}{Lemma}
\newtheorem{Exa}{Example}
\newtheorem{Rem}{Remark}
\newtheorem{Cor}{Corollary}
\renewcommand{\labelitemi}{$\bullet$}

\newcommand{\ist}{\hspace*{.3mm}}
\newcommand{\rmv}{\hspace*{-.3mm}}
\newcommand{\iist}{\hspace*{1mm}}
\newcommand{\rrmv}{\hspace*{-1mm}}
\newcommand{\nn}{\nonumber}
\newcommand{\sist}{\hspace*{.15mm}}
\newcommand{\trans}{^\mathrm{T}}

\title{Normalizing Flow for\\ Neural-Enhanced Track-Before-Detect}
\author{\IEEEauthorblockN{Shaoxiu Wei, Bhaskar Rao, and Florian Meyer}\vspace{.8mm}
\IEEEauthorblockA{University of California San Diego, La Jolla, CA (\{shwei,brao,flmeyer\}@ucsd.edu)\vspace{1.5mm}}
}

\providecommand{\bu}{\textcolor{blue}}
\providecommand{\rd}{\textcolor{red}}
\maketitle

\begin{abstract}
Multiobject tracking (MOT) is a fundamental capability in autonomous navigation, radar and sonar systems, as well as 6G integrated sensing-and-communication (ISAC) networks. In this paper, we propose a neural-enhanced track-before-detect (TkBD) framework that operates directly on complex-valued raw signals and learns a latent representation of unmodeled propagation effects during runtime. We introduce an environment-dependent latent intensity variable that characterizes the signal power of each object in the shared propagation environment. The probability density function of this intensity variable is represented by a normalizing flow and trained online in an unsupervised manner within the sequential Bayesian inference procedure. The learning objective is derived from a variational Bayes formulation by maximizing the evidence lower bound (ELBO), thereby accounting for the uncertainties inherent in the Bayesian TkBD problem. To enable scalable inference, we make use of a recently introduced Bayesian TkBD framework that combines belief propagation, which avoids high-dimensional operations, with sequential Monte Carlo methods for efficient low-dimensional updates.
\end{abstract}
\begin{IEEEkeywords}
	Multiobject tracking, Bayesian estimation, belief propagation, track-before-detect, and normalizing flow.
\end{IEEEkeywords}

\section{Introduction}
\label{sec:intro}

The ability to perceive, interpret, and navigate dynamic environments is a key requirement in autonomous navigation, radar and sonar systems, and emerging 6G integrated sensing-and-communication (ISAC) networks. Multiobject tracking (MOT) \cite{BarWilTia:B11,Mah:B07,Bla:J04,Wil:J15,MeyWil:J21} addresses this task by estimating the time-varying states of an unknown and time-varying number of objects from sensor measurements. However, reliable MOT remains challenging when objects are weak, closely spaced, intermittently visible, or affected by complex propagation phenomena. In addition, object appearance and disappearance are typically unknown, requiring robust track management for object initialization and termination.

Most conventional MOT methods follow a detect-then-track (DTT) paradigm. In a first stage, a nonlinear detector preprocesses the raw sensor data to reduce data rate and computational complexity. In a second stage, the resulting detections, or ``point measurements,'' are passed to a tracking algorithm \cite{BarWilTia:B11,Mah:B07}. While effective when detections are reliable, this modular design can suffer from measurement-origin uncertainty and irreversible information loss in the detection stage. These limitations are especially severe in low-signal-to-noise-ratio (SNR) and low-observability regimes. Classical Bayesian MOT methods address association ambiguity by means of multiple hypothesis tracking \cite{Rei:J79,Bla:J04}, joint probabilistic data association \cite{BloBl:95,HabThaThaMalKir:J13}, random finite set methods \cite{LunGraOrg:J13,ReuVoVoDie:J14,VoMa:J06,GarWilGraSve:J18}, and belief propagation (BP)-based inference \cite{MeyBraWilHla:15,MeyKroWilLauHlaBraWin:J18,LiaMey:J23,JanMeySnyWigBauHil:J23,WeiLiaMey:J26}. Nevertheless, the frontend detection stage can discard weak but valuable object information before Bayesian tracking is performed.

Track-before-detect (TkBD) methods aim to overcome this limitation by avoiding a hard detection step and operating directly on received signals \cite{Bar:J85,MoySpaLam:J11}. Conventional TkBD methods typically apply a linear, physics-motivated mapping that transforms raw signals into data cells, such as range--Doppler, angle--range, or delay--Doppler--angle cells, which are then used as measurements for Bayesian filtering. By avoiding early thresholding, TkBD can preserve weak object evidence and improve tracking performance when returns are faint or overlapping.

\begin{figure}[!t]
    \centering
    \includegraphics[scale=1.45]{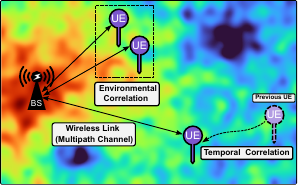}
    \caption{Multiobject tracking in a wireless communication network with a base station (BS) and user equipment (UE). Raw received data are used directly as measurements, enabling unmodeled propagation effects to be learned and exploited for improved tracking performance\vspace{-5mm}.}
    \label{Scene}
\end{figure}

These advantages are particularly important in 6G ISAC networks \cite{DelBelBer:J21,LiuMasPet:J22}, where infrastructure nodes are expected to localize and track user equipment (UE) as well as environmental reflectors. In such systems, complex propagation effects, including multipath and scattering, can provide useful information for localization and tracking \cite{KadBehSorWelAmiYerYoo:22,LiaLeiMey:J25}. However, traditional model-based DTT and TkBD approaches often rely on simplified parametric propagation models \cite{LiaMey:J26,VoVo:J13,MeyKroWilLauHlaBraWin:J18,DavGae:B18,BuzLopVenFer:J08}. These models can be brittle in realistic environments, whereas purely learning-based methods usually require large labeled datasets and may provide limited interpretability or generalization \cite{ShlEld:J23,WeiLiaMey:J26}. This motivates hybrid methods that preserve physically meaningful measurement models while learning unmodeled propagation effects from data.

In this paper, we propose a neural-enhanced TkBD framework for MOT that operates directly on complex-valued raw signals. The key idea is that raw measurements contain information about unmodeled propagation effects that is typically discarded when observations are compressed into hand-crafted low-dimensional features. The proposed method therefore infers a latent probabilistic representation of these propagation effects and exploits this representation to improve tracking performance. We introduce an environment-dependent latent intensity variable $\gamma$ that characterizes the signal power of each object in the shared propagation environment. The prior intensity probability density function (PDF) $p(\gamma;\mathbf{c})$ is represented by a normalizing flow \cite{KobPriBru:J21}, where $\mathbf{c}$ contains geometry-dependent features, and is trained online in an unsupervised manner within the sequential Bayesian inference procedure. The learning objective is obtained from a variational Bayes formulation by maximizing the evidence lower bound (ELBO), thereby accounting for the uncertainties of the Bayesian TkBD problem. Scalable inference is achieved by combining BP, which avoids high-dimensional multiobject computations, with sequential Monte Carlo (SMC) methods for efficient low-dimensional updates.

The main contributions of this paper are as follows.
\begin{itemize}
    \item We introduce an environment-dependent latent intensity variable $\gamma$ that enables TkBD to learn and exploit unmodeled propagation effects\vspace{1.5mm}.
    \item We model the prior intensity PDF $p(\gamma;\mathbf{c})$ using a normalizing flow that is trained online and without supervision within the sequential Bayesian inference procedure\vspace{1.5mm}.
    \item We demonstrate numerically that the proposed neural-enhanced TkBD method can significantly improve MOT performance in a radar tracking scenario\vspace {0mm}.
\end{itemize}

\section{Joint Estimation and Learning Framework}
\label{sec:framework}

In this section, we introduce the statistical model of the considered neural-enhanced TkBD problem and formulate the corresponding joint estimation and learning task. We consider a multi-sensor MOT problem. Measurements are available at the discrete time steps
$k \in \{1,\ldots,K\}$. The number of objects is unknown and time-varying. To facilitate Bayesian estimation, at each time step $k$, we consider $N_k$ potential objects (POs) indexed by $n \in \{1,\ldots,N_k\}$. The kinematic state of PO with index $n$ is denoted by
$\V{x}_{k,n} \in \mathbb{R}^{d_x}$ with prior PDF $p(\V{x}_{k,n})$.

Object existence is modeled by a Bernoulli random variable
$r_{k,n} \in \{0,1\}$ with prior probability mass function (PMF) $p(r_{k,n})$, where $r_{k,n}=1$ indicates that
PO $n$ exists and contributes to the received measurements at time $k$. We consider $I$ sensors indexed by $i \in \{1,\ldots,I\}$.
At time $k$, sensor $i$ produces complex-valued raw data
$\V{z}_k^{(i)} \in \mathbb{C}^{m_i}$ after vectorization, where $m_i$ denotes
the channel dimension. Given a kinematic PO state $\V{x}$, sensor $i$
is associated with a known steering vector
$\V{a}^{(i)}(\V{x}) \in \mathbb{C}^{m_i}$ determined by sensor geometry, system calibration, and the propagation environment.

To describe propagation and clutter effects that are not fully explained by geometry alone, we introduce a deterministic context vector $\V{c}_{k,n}^{(i)}$. The vector $\V{c}_{k,n}^{(i)}$ may contain, for example, map information, sensor metadata, or object location estimates, and serves as a hyperparameter of the intensity PDF represented by a normalizing flow \cite{KobPriBru:J21}.

\subsection{Statistical Model and BP-Based Estimation}

The basic idea is to establish a physical mapping between object states and raw measurements. This mapping is governed by two components: (i) the steering vector, which is determined by sensor geometry and dominant modeled propagation effects, and (ii) the intensity, which represents the strength of the object contribution to the superpositional sensor measurement. While the steering vector follows a prescribed model, the intensity can depend on target scattering strength, multipath propagation, and other propagation effects that are difficult to capture by a simple parametric model. This motivates a measurement model that combines physically grounded steering vectors with a learned prior model for signal intensity.

For each sensor $i$, the measurement model is given as the superposition of existing objects plus noise, i.e.\vspace{-.5mm},
\begin{equation}
	\V{z}_k^{(i)} = \sum_{n=1}^{N_k} r_{k,n}\,\rho_{k,n}^{(i)}\, \V{a}^{(i)}(\V{x}_{k,n}) + \V{\epsilon}_k^{(i)}. \nn
	\vspace{-.5mm}
\end{equation}
We model the complex signal amplitude, $\rho_{k,n}^{(i)} \in \mathbb{C}$, of object $n$ at time $k$ for sensor $i$ by the conditional\vspace{.7mm} PDF
\begin{equation}
	\rho_{k,n}^{(i)} \mid \gamma_{k,n}^{(i)} \sim \mathcal{CN} \big(0,\gamma_{k,n}^{(i)}\big) \nn
\end{equation}
where $\gamma_{k,n}^{(i)} > 0$ is the object-dependent intensity at sensor $i$. Furthermore, sensor noise is modeled as
\begin{equation}
	\V{\epsilon}_k^{(i)} \mid \eta_k^{(i)} \sim \mathcal{CN} \big(0,\eta_k^{(i)} \M{I}_{m_i}\big) \nn
\end{equation}
with unknown noise power $\eta_k^{(i)} > 0$. Based on these PDFs, the measurement can be represented as a zero-mean complex Gaussian, i.e.,
\begin{equation}
	\V{z}_k^{(i)} \mid r_k, \V{x}_k, \gamma_k^{(i)}, \eta_k^{(i)} \nn
	  \sim \mathcal{CN}\!\big(0,\, \M{C}_k^{(i)}\big)
\end{equation}
with\vspace{.5mm} covariance matrix
\begin{equation}
	\M{C}_k^{(i)} =
	\sum_{n=1}^{N_k} r_{k,n}\,\gamma_{k,n}^{(i)}\, \V{a}^{(i)}(\V{x}_{k,n}) \ist \V{a}^{(i)}(\V{x}_{k,n})^{\mathrm{H}}
	+ \eta_k^{(i)} \M{I}_{m_i}. \nn
\end{equation}

The key novel modeling component is the learnable prior PDF for the intensity
$\gamma_{k,n}^{(i)}$, i.e.\vspace{-.7mm},
\begin{equation}
	\gamma_{k,n}^{(i)} \sim p_{\V{\theta}}\!\big(\gamma_{k,n}^{(i)} ; \V{c}_{k,n}^{(i)}\big).  \nn
\end{equation}
This model incorporates the context $\V{c}_{k,n}^{(i)}$ and aims at capturing unmodeled propagation effects. The learnable parameter $\V{\theta}$ defines the functional form of this PDF. We adopt standard Markovian dynamics for existence, kinematics, and noise power denoted as $p(r_{k,n} \ist|\ist r_{k-1,n})$, $p(\V{x}_{k,n} \ist|\ist  \V{x}_{k-1,n})$, and $p(\eta_k^{(i)} \ist|\ist  \eta_{k-1}^{(i)})$. Next, we collect all random variables at time $k$ in a joint vector, i.e.\vspace{-.7mm}, 
\begin{equation}
	\V{h}_k =
	\Big(
	r_{k,1:N_k},\,
	\V{x}_{k,1:N_k},\,
	\{\gamma_{k,1:N_k}^{(i)}\}_{i=1}^I,\,
	\{\eta_k^{(i)}\}_{i=1}^I
	\Big). \nn
\end{equation}
Conditioned on $\V{h}_k$, measurements are independent across sensors. The likelihood function can thus be written as the product of complex Gaussian PDFs, i.e.,
\begin{equation}
	p(\V{z}_k \mid \V{h}_k)
	= \prod_{i=1}^I \mathcal{CN}(\V{z}_k^{(i)}; \V{0}, \M{C}_k^{(i)}). \nn
\end{equation}
Let us introduce the joint vector $\V{h}_{1:K} := \{\V{h}_1,\ldots,\V{h}_K\}$. The overall PDF across all sensors, objects, and time steps can be factorized \vspace{-1mm} as \cite{LiaMey:J26}
\begin{equation}
	p_{\V{\theta}}(\V{z}_{1:K}, \V{h}_{1:K})
	= p(\V{h}_1) \ist p(\V{z}_1 \ist|\ist \V{h}_1)   \prod_{k=2}^K p(\V{z}_k \ist|\ist \V{h}_k) \ist p_{\V{\theta}}(\V{h}_k \ist|\ist \V{h}_{k-1}) 
	\label{eq:joint}
\end{equation}
where the transition term expands as
\begin{align}
	p_{\V{\theta}}(\V{h}_k\mid \V{h}_{k-1})
	&=
   \Bigg( \prod_{i=1}^I
	p(\eta_k^{(i)}\mid \eta_{k-1}^{(i)}) \rmv\rmv \Bigg) \rmv \prod_{n=1}^{N_k} p(r_{k,n}\mid r_{k-1,n}) \notag\\[0mm]
	&\times\, p(\V{x}_{k,n}\mid\V{x}_{k-1,n})
	  \prod_{i=1}^I
	p_{\V{\theta}}(\gamma_{k,n}^{(i)} ; \V{c}_{k,n}^{(i)})\rmv.	\label{eq:transition_full} \\[-6mm]
	\nn
\end{align}
Note that new object states are introduced as in \cite{MeyKroWilLauHlaBraWin:J18,LiaMey:J26}.

The declaration of object existence and the estimation of object states at time $k$, rely on marginal posterior probability mass functions $p(r_{k,n} \ist|\ist \V{z}_{1:k})$ and PDFs $p(\V{x}_{k,n} \ist|\ist \V{z}_{1:k})$ for all $n \in \{1,\dots,N_k\}$ (see \cite{MeyKroWilLauHlaBraWin:J18}) for details. However, direct computation of these marginals from the posterior PDF $p_{\V{\theta}}(\V{h}_{1:k} \ist | \ist \V{z}_{1:k})$ is intractable due to the high-dimensionality of $\V{h}_{1:k}$. 

Following \cite{LiaMey:J26}, we perform loopy BP based on the factor graph defined by the factorization in \eqref{eq:joint}--\eqref{eq:transition_full}. In particular, at each time step $k$, BP provides particle-based representations of approximate marginal posterior PDFs/PMFs, i.e.,
\begin{align*}
   q(r_{k,n}) &\approx p(r_{k,n} \ist| \ist  \V{z}_{1:k}) \\[2mm]
   q(\V{x}_{k,n}  \ist| \ist  r_{k,n}=1) &\approx p(\V{x}_{k,n}  \ist| \ist  r_{k,n}=1,\V{z}_{1:k})\\[2mm]
   q(\gamma_{k,n}^{(i)})&\approx p(\gamma_{k,n}^{(i)}  \ist| \ist  \V{z}_{1:k})\\[2mm]
   q(\eta_k^{(i)})&\approx p(\eta_k^{(i)}  \ist| \ist  \V{z}_{1:k}). \nn
\end{align*}
For example, at time $k$, a representation of the approximate marginal posterior PDFs for the object state with index $n$ and its intensity, can be written\vspace{-.5mm} as
\[
q(\V{x}_{k,n}  \ist| \ist  r_{k,n}=1) = \left\{\big(\V{x}_{k,n}^{(j)},\omega^{(j)}_{k,n}\big)\right\}_{j=1}^{J_{\V{x}}}
\]
and
\begin{equation}
q(\gamma_{k,n}^{(i)}) = \left\{\big(\gamma_{k,n}^{(i,j)},\omega_{k,n}^{(i,j)}\big)\right\}_{j=1}^{J_\gamma}. \label{eq:gammaUpdate}
\end{equation}
A similar particle-based representation is obtained for the background noise power. 

Object existence is determined by comparing $q(r_{k,n}=1)$ with an object declaration threshold. Finally, the state of objects that have been declared to exist is estimated using a particle-based approximation of the minimum-mean-square-error (MMSE) estimator \cite{Kay:B93}, i.e.\vspace{-.5mm},
\begin{align}
	\hat{\mathbf{x}}_{k,n}^{\mathrm{MMSE}} 
	&= \sum_{j=1}^{J_{\V{x}}} \mathbf{x}^{(j)}_{k,n} \ist \omega^{(j)}_{k,n}. \nn \\[-5mm]
	\nn
\end{align}

\subsection{ELBO-Based Learning Objective}

\begin{figure}[t]
	\centering
	\includegraphics[width=3.4in]{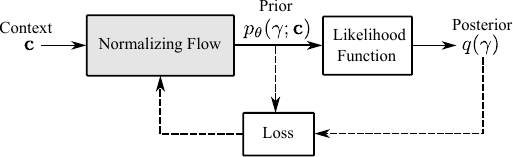}
	\caption{Block diagram of the normalizing-flow-based module used to learn the prior intensity PDF $p(\gamma;\mathbf{c})$ online\vspace{0mm}.}
	\label{VB1}
\end{figure}

We next develop the learning objective for determining the parameter $\V{\theta}$ of the prior PDF for the intensity. To this end, we consider an approximate posterior PDF $q(\V{h}_{1:K})$. For any approximation $q(\V{h}_{1:K})$\vspace{1mm}, we can write \cite{Bis:B06}
\begin{align}
	\log p_{\V{\theta}}(\V{z}_{1:K})
	&= \mathcal{L}(\V{\theta},q) + \mathrm{KL} \big(q(\V{h}_{1:K})||p_{\V{\theta}}(\V{h}_{1:K}|\V{z}_{1:K})\big).\nn\\[-2.5mm]
	\nn
\end{align}
Here, we introduced the nonnegative Kullback-Leibler (KL) divergence and the quantity $\mathcal{L}(\V{\theta},q)$. It can easily be verified that $\mathcal{L}(\V{\theta},q)$ is a lower bound of $\log p_{\V{\theta}}(\V{z}_{1:K})$. We use this evidence lower bound (ELBO) as a learning objective for the training of $p_{\V{\theta}}\rmv\big(\gamma_{k,n}^{(i)} ; \V{c}_{k,n}^{(i)}\big)$. The ELBO can be further decomposed as
\begin{align}
	\mathcal{L}(\V{\theta},q)
	&= \mathbb{E}_q\big[\log p_{\V{\theta}}(\V{z}_{1:K},\V{h}_{1:K})\big] - \mathbb{E}_q\big[\log q(\V{h}_{1:K})\big]\nonumber\\[2.5mm]
	&= \sum_{k=1}^K \underbrace{\mathbb{E}_q\!\big[\log p(\V{z}_k| \V{h}_k)\big]}_{\text{measurement model}}
	+ \sum_{k=2}^K \underbrace{\mathbb{E}_q\!\big[\log p_{\V{\theta}}(\V{h}_k\mid \V{h}_{k-1})\big]}_{\text{state-transition}} \notag\\[2.5mm]
	&+ \underbrace{\mathbb{E}_q\!\big[\log p(\V{h}_1)\big]}_{\text{initial prior}}
	- \mathbb{E}_q\!\big[\log q(\V{h}_{1:K})\big]. \nn\\[-3.7mm]
	\nn
\end{align}
At time $k$, the term of the ELBO that depends on the parameters $\V{\theta}$, reads
\begin{equation}
		\mathcal{L}_{k}(\V{\theta},q)
		= \rmv\rmv \sum^{k}_{k'=2} \sum^{N_{k'}}_{n=1} \sum^{I}_{i=1}
		\mathbb{E}_{q(\gamma^{(i)}_{k',n})} \rmv\rmv \big[\log p_{\V{\theta}}(\gamma^{(i)}_{k',n};\,\V{c}^{(i)}_{k',n})\big]. 
	\label{eq:Lgamma}
\end{equation}
Thus, at time $k$, the value of $\V{\theta}$ that maximizes the ELBO can be obtained\vspace{-.8mm} as
\begin{align}
	\V{\theta}^\star =&~ \arg\max_{\V{\theta}}\;\mathcal{L}_{k}(\V{\theta},q). \nn
\end{align}

Variational inference \cite{Bis:B06} is a common approach for efficient estimation when the PDF
$p_{\V{\theta}}(\V{h}_{1:K} \ist | \ist \V{z}_{1:K})$ is intractable. It introduces a simpler variational PDF $q(\V{h}_{1:K} )$ to approximate the posterior PDF $p_{\V{\theta}}(\V{h}_{1:K} \ist | \ist \V{z}_{1:K})$ and alternates between updating $q(\V{h}_{1:K} )$ for fixed $\V{\theta}$ and updating $\V{\theta}$ by maximizing the ELBO. In the proposed approach, BP directly provides a particle-based representation of $q(\gamma^{(i)}_{k,n})$ as in \eqref{eq:gammaUpdate} as will be discussed at the end of this section. Thus, we treat $q(\gamma^{(i)}_{k,n})$ as given and do not perform any additional variational update to compute $q(\gamma^{(i)}_{k,n})$.  The expectation in \eqref{eq:Lgamma} is computed using Monte Carlo integration based on the particle representation
of $q(\gamma_{k,n}^{(i)})$ in \eqref{eq:gammaUpdate}. Note that this particle representation is different for each time step $k$, object $n$, and sensor $i$. 

\subsection{The Learnable Prior PDF $p_{\V{\theta}}\!\big(\gamma_{k,n}^{(i)} ; \V{c}_{k,n}^{(i)}\big)$}

We next describe how to represent and evaluate the learnable prior PDF $p_{\V{\theta}}\!\big(\gamma_{k,n}^{(i)} ; \V{c}_{k,n}^{(i)}\big)$ using a context-dependent normalizing flow \cite{KobPriBru:J21}. Since it is convenient to work in an unconstrained domain but $\gamma^{(i)}_{k,n} > 0$, we introduce a mapping
\begin{equation}
	\gamma^{(i)}_{k,n} = g(\mu^{(i)}_{k,n}), \qquad  \mu^{(i)}_{k,n} \in \mathbb{R}. \nn
\end{equation}
For simplicity, we omit the time, object, and sensor index $(k,n,i)$ in the following discussions.

Let $p_0(\varepsilon)$ be the initial PDF
on $\mathbb{R}$, e.g.\ $\varepsilon \sim \mathcal{N}(0,1)$.
For a given context vector $\V{c}$, we define the flow as a sequence of $L$ invertible,
monotone transformations, i.e.,
\begin{equation}
	y_0 = \varepsilon,\quad
	y_1 = f_{\V{\theta}_1}(y_0; \V{c}),\quad
	\dots,\quad
	y_{L} = f_{\V{\theta}_{L}}(y_{L-1}; \V{c}). \nn
\end{equation}
Here, $f_{\V{\theta}_\ell}(\cdot; \V{c})$ is a differentiable
bijection $\mathbb{R} \to \mathbb{R}$ for fixed context $\V{c}$. The result of this sequence of transformations can be written \vspace{-1mm} as
\begin{equation}
	\mu = y_{L} = T_{\V{\theta}}(\varepsilon;\V{c}) \nn
\end{equation}
where $T_{\V{\theta}}(\varepsilon;\V{c})$ is the joint mapping and $\V{\theta} = (\V{\theta}_1,\dots,\V{\theta}_{L})$ are the flow parameters.

Using the change-of-variables formula \cite{PapPil:02}, the log-density of
$\mu$ as a function of $\V{c}$, can be evaluated \vspace{-1mm} as 
\begin{equation}
\label{eq:pdf_compute_midlayer}
	\log p_{\V{\theta}}(\mu ; \V{c})
	= \log p_0(\varepsilon) -
	\sum_{\ell=1}^{L}
	\log \left|
	\frac{\partial f_{\V{\theta}_\ell}(y_{\ell-1}; \V{c})}%
	{\partial y_{\ell-1}}
	\right|
\end{equation}
where $\varepsilon$ and $y_{\ell}$ are obtained by inverting the individual bijections
for a given $(\mu, \V{c}_k)$. Finally,  by again using the change-of-variables formula \cite{PapPil:02}, we can obtain the log-density of the learnable prior PDF\vspace{-.5mm} as 
\begin{equation}
\label{eq:pdf_compute_lastlayer}
	\log p_{\V{\theta}}(\gamma ; \V{c})
	=
	\log p_{\V{\theta}}(\mu ; \V{c})
	- \log \left|
	\frac{\partial \gamma}%
	{\partial \mu}
	\right|.
\end{equation}

Once the learnable prior PDF $p_{\V{\theta}}(\gamma ; \V{c})$ has been
learned, it can be used to generate prior particles for the intensity
at any time step $k$, given the corresponding context $\V{c}$.
Sampling from and evaluation of $p_{\V{\theta}}(\gamma ; \V{c})$ can be summarized as follows.
\begin{enumerate}
	\item For each particle $j \in \{1,\dots,J_{\gamma}\}\vspace{.3mm}$.
	\begin{enumerate}
		\item Draw a base sample $\varepsilon^{(j)} \sim p_0(\varepsilon)$
		(e.g.\ $\mathcal{N}(0,1)$)\vspace{.3mm}.
		\item Propagate the sample using the flow, i.e.,
		\begin{equation}
			y_0^{(j)} = \varepsilon^{(j)},\quad
			y_\ell^{(j)} = f_{\theta_\ell}\big(y_{\ell-1}^{(j)}; \V{c}\big),
			\quad \ell = 1,\dots,L, \nn
		\end{equation}
		and set $\mu^{(j)} = y_{L}^{(j)}$\vspace{.3mm}.
		\item Transform the sample back to intensity space, i.e.,
		\begin{equation}
			\gamma^{(j)} = g(\mu^{(j)}) > 0. \nn
		\end{equation}
		\item Evaluate the value of $\log p_{\V{\theta}}(\gamma^{(j)} ; \V{c})$ by using \eqref{eq:pdf_compute_midlayer} and \eqref{eq:pdf_compute_lastlayer}\vspace{.3mm}.
	\end{enumerate}
\end{enumerate}

Note that in the update step of BP \cite{LiaMey:J26}\vspace{-.5mm}, weights $\omega^{(i,j)}_{k,n}$ are computed for each prior particle $\gamma^{(i,j)}_{k,n}$. The resulting set of prior particles and their weights in \eqref{eq:gammaUpdate}, represents the approximation of the marginal posterior PDF $q(\gamma_{k,n}^{(i)})$. This set of particles can be directly used to evaluate the learning objective in \eqref{eq:Lgamma}, i.e.\vspace{-.5mm},
\begin{align}
	\mathcal{L}_{k}(\V{\theta},q) &= \sum^{J_{\gamma}}_{j=1}  \sum^{k}_{k'=2} \sum^{N_{k'}}_{n=1} \sum^{I}_{i=1} \omega^{(i,j)}_{k',n}  \log p_{\V{\theta}}\big(\gamma^{(i,j)}_{k',n};\V{c}^{(i)}_{k',n}\big). \nn\\[-6mm]
	\nn
\end{align}
Note that evaluating the learning objective relies on the evaluation of $\log p_{\V{\theta}}\!\big(\gamma_{k,n}^{(i)} ; \V{c}_{k,n}^{(i)}\big)$ as detailed in Step d).
\vspace{0mm}

\section{Numerical Analysis}
In this section, we evaluate the performance of our proposed
method based on synthetic data.

\subsection{Synthetic Radar Tracking Scenario}
We consider a two-dimensional (2-D) multi-sensor radar system with $I$ sensors equipped with a uniform linear array.
Sensors are located at positions $\mathbf{p}^{(i)}$, $i\in\{1,\ldots,I\}$. A linear chirp signal is transmitted and reflections from objects are
received. This type of system is referred to as linear frequency-modulated continuous-wave (LFMCW) radar. At
each time step $k$, the radar system demodulates, downconverts,
and discretizes the analog received signal to obtain the baseband signal $z^{(i)}_{k,j}$. The steering vector is given by
\begin{equation}
	\mathbf{a}^{(i)}(\mathbf{x})
	= \mathrm{vec}\Big(\mathbf{a}^{(i)}_{f}\otimes\mathbf{a}^{(i)}_{s} \otimes\mathbf{a}^{(i)}_{\vartheta} \Big) \nn
\end{equation}
where $\otimes$ represents the Kronecker product. The fast time (delay), slow time (Doppler), and angular steering vector components are defined as
\begin{align*}
	\mathbf{a}^{(i)}_{f}
	&=
	\Big[1~~ e^{j4\pi \mu T_s d^{(i)}/c}~~ \cdots~~ e^{j4\pi (M_f-1)\mu T_s d^{(i)}/c}\Big]^{\mathrm{T}}
	\\
	\mathbf{a}^{(i)}_{s}
	&=
	\Big[1~~ e^{j4\pi T_c v^{(i)}/\lambda}~~ \cdots~~ e^{j4\pi (M_s-1)  T_c v^{(i)}/\lambda}\Big]^{\mathrm{T}}
	\\
	\mathbf{a}^{(i)}_{\vartheta}
	&=
	\Big[1~~ e^{j\pi \sin(\vartheta^{(i)})}~~ \cdots~~ e^{j\pi (M_\vartheta-1)\sin(\vartheta^{(i)})}\Big]^{\mathrm{T}}\rmv\rmv.
\end{align*}
Here $d^{(i)}$ denotes the distance between object and sensor $i$ and $v^{(i)}$ denotes the radial velocity of object to sensor $i$, while
$\vartheta^{(i)}$ is the angle of arrival (AOA) relative to the array orientation.
The orientation of both arrays is $\frac{\pi}{4}$ and the antenna spacing is half of the wavelength. The parameters $c$, $\mu$, $\lambda$, $T_s$ and $T_c$ represent the speed of light,
chirp rate, wavelength, sampling interval, and pulse repetition interval (PRI), respectively. The total measurement size is $M=M_fM_s \ist M_\vartheta$, combining $M_f$ fast time samples, $M_s$ slow time samples,
and $M_\vartheta$ array elements. Here, $\mathrm{vec}(\cdot)$ is the vectorization operation. For data generation, the power of the
complex amplitude $\varrho^{(i)}$ is characterized by the radar-range equation
\begin{equation}
	\label{eq:radar_eq}\big|\varrho^{(i)}\big|^2
	=
	\frac{c^2}{f_c^{2}(4\pi)^{3}\big(d^{(i)}\big)^{4}}. 
\end{equation}
where $f_c=77~\mathrm{GHz}$ is the carrier frequency. The phase of $\varrho^{(i)}$ is uniformly distributed over $[0,2\pi)$.

In this experiment, there are $I=2$ low-resolution radar sensors at positions $\mathbf{p}^{(1)}=[0~~0]^{\mathrm{T}}$ and
$\mathbf{p}^{(2)}=[50~~50]^{\mathrm{T}}$. We set the chirp rate to $\mu=8~\mathrm{MHz}/\mu\mathrm{s}$, the sampling interval to
$T_s=0.2~\mu\mathrm{s}$, the PRI to $T_c = 40~\mu\mathrm{s}$, the number of fast time samples $M_f=20$, the number of slow time samples $M_s=10$, and the number of array elements to $M_\vartheta=10$. 

A total of $80$ time steps and five objects were simulated. Objects are moving within a region of interest (ROI) defined by the intervals $x = [0, 50]$m and $y = [0, 50]$m. Parameters describing the simulated object tracks are provided in Table \ref{object_info}.
\begin{table}[!t]
	\centering
	\caption{Parameters describing simulated object tracks.}
	\label{object_info}
	\resizebox{\linewidth}{!}
	{\begin{tabular}{c|c|c} 
			\hline
			Object Index& Initial State (Pos. + Vel.) &Birth \& Death Time\\ 
			\hline
			1&[10,25,0.5,0]&[2, 72]\\   
			2&[25,10,0,0.5]&[5, 75]\\ 
			3&Random&[10, 70]\\ 
			4&Random&[18, 78]\\ 
			5&Random&[25, 75]\\ 
			\hline
		\end{tabular}
	}
\end{table}
We conduct $100$ simulation runs. For data generation, the noise power is set such that the ``input'' SNR is equal to
$10\log_{10}\big(|\varrho^{(i)}_{k,\ell}|^2/\eta^{(i)}_{k}\big)\rmv=\rmv-3$~dB for objects at the center of the ROI. The kinematic state $\mathbf{x}_{k,n}=[\mathbf{p}_{k,n}^{\mathrm{T}},~\mathbf{v}_{k,n}^{\mathrm{T}}]^{\mathrm{T}}$ of PO
$n\in\{1,\ldots,N_k\}$ contains its 2-D position, $\mathbf{p}_{k,n}$, and 2-D velocity, $\mathbf{v}_{k,n}$. A constant velocity model
$\mathbf{x}_{k,n}=\mathbf{F}\mathbf{x}_{k-1,n}+\mathbf{W}\mathbf{q}_{k}$ is considered (see \cite[Sec. 6.3.2]{ShaKirLi:B02} for details). The context vector used as an input for the learnable prior PDF includes estimated distances and angles between objects and sensors.

\subsection{Performance Evaluation}
MOT performance is evaluated using the generalized optimal subpattern assignment (GOSPA) metric \cite{RahGarSve:17} with cutoff parameter $c=5$ and order $p=1$. We compare the proposed method with three representative baselines:
\begin{figure}[!t]
	\centering
	\includegraphics[width=3.5in]{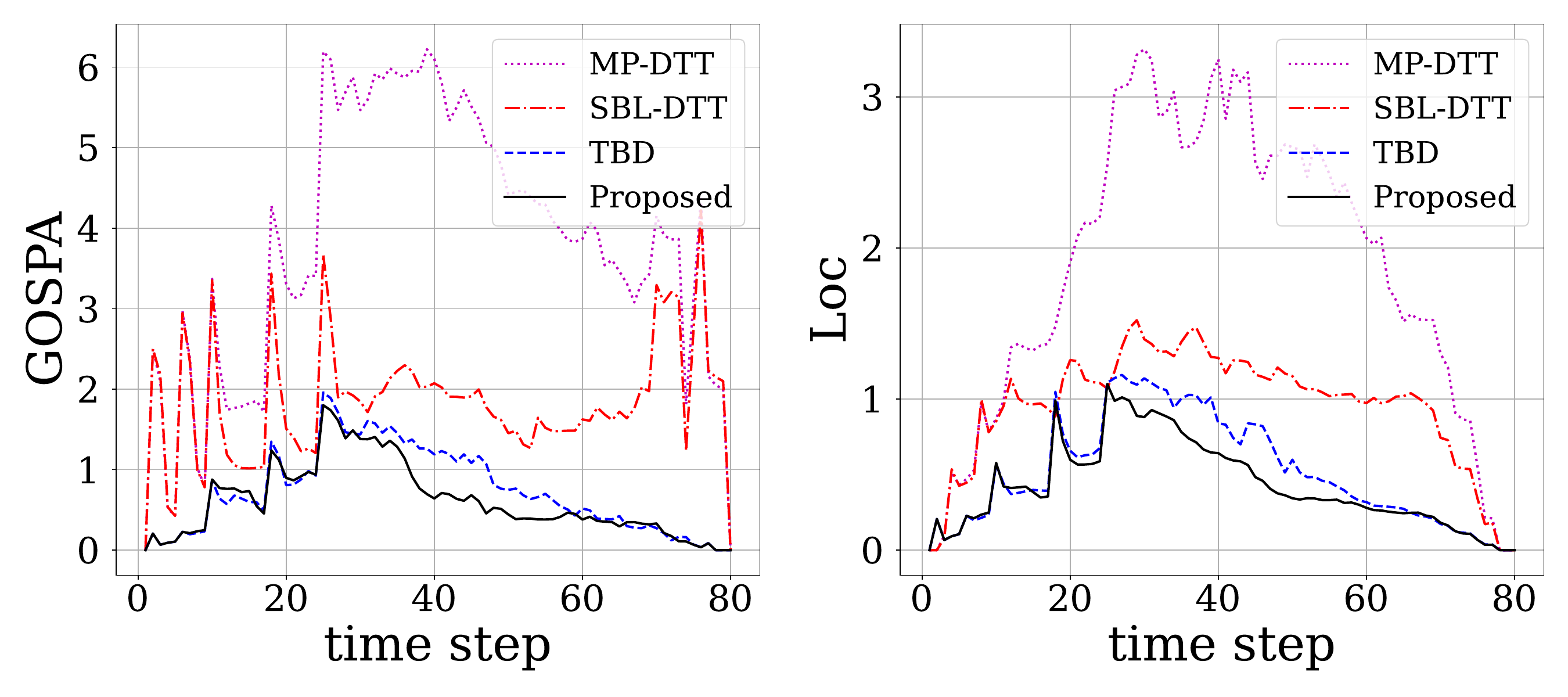}
	\caption{Total GOSPA error and localization-error component obtained by the four considered tracking methods over $100$ simulation runs\vspace{0mm}.}
	\label{GOSPA}
	\vspace{-3mm}
\end{figure}

\begin{itemize}
	\item \textbf{MP-DTT} uses detections provided by matching pursuit \cite{MalZha:J93} within BP-based DTT \cite{MeyBraWilHla:J17}; $30000$ particles are used to represent the kinematic states\vspace{.5mm}.
    \item \textbf{SBL-DTT} uses detections provided by sparse Bayesian learning \cite{WipRao:J07,GerMecChrXenNan:J16} within BP-based DTT \cite{MeyBraWilHla:J17}; $30000$ particles are used to represent the kinematic states.
	\item \textbf{TkBD} performs BP-based TkBD with an intensity prior that follows a random-walk model \cite{LiaMey:J26}; $10000$ particles are used to represent kinematic states, $500$ particles to represent intensity, and $500$ particles to represent measurement-noise variances\vspace{.5mm}.
	\item \textbf{Proposed TkBD} performs BP-based TkBD with a learnable intensity prior provided by a normalizing flow; $10000$ particles are used to represent kinematic states, $500$ particles to represent intensity, and $500$ particles to represent measurement-noise variances.
\end{itemize}

\par Our numerical performance evaluation is summarized in Fig.~\ref{GOSPA}. Among the four methods, the proposed approach consistently achieves the lowest total GOSPA and the lowest localization error (LOC) over time. In contrast, the DTT baselines exhibit noticeably larger errors, which can be attributed to incorrect decisions during the detection stage (e.g., missed detections and false alarms) as well as uncertainty in data association. The conventional TkBD baseline outperforms DTT approaches. However, its random-walk intensity model cannot fully capture the data-generating process. By learning the prior PDF of the intensity online, the proposed TkBD approach can outperform all reference methods.

\begin{figure}[t]
	\centering
	
	\begin{subfigure}[b]{0.49\linewidth}
		\centering
		\includegraphics[width=1.7in]{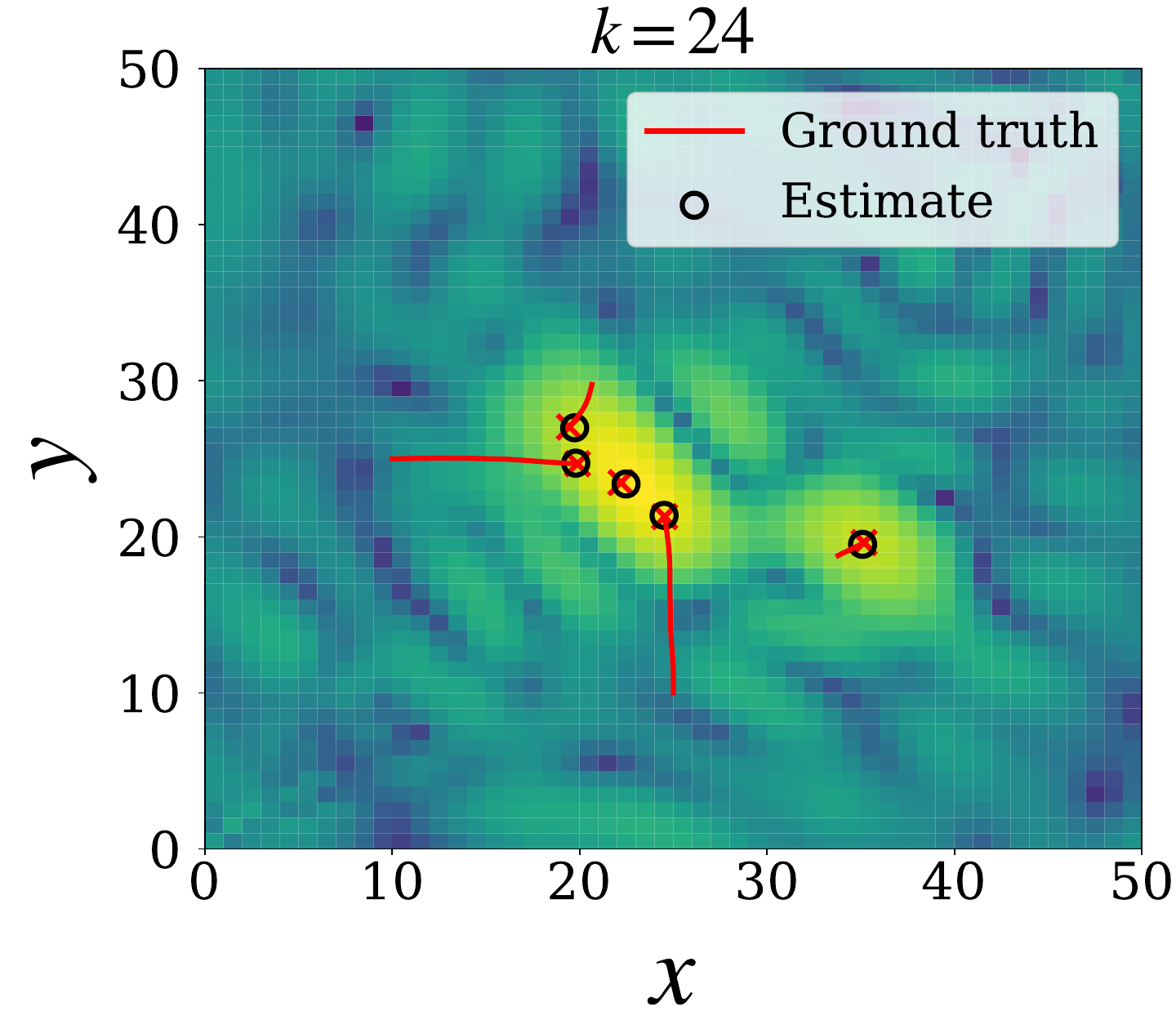}
	\end{subfigure}
	\hfill
	\begin{subfigure}[b]{0.49\linewidth}
		\centering
		\includegraphics[width=1.7in]{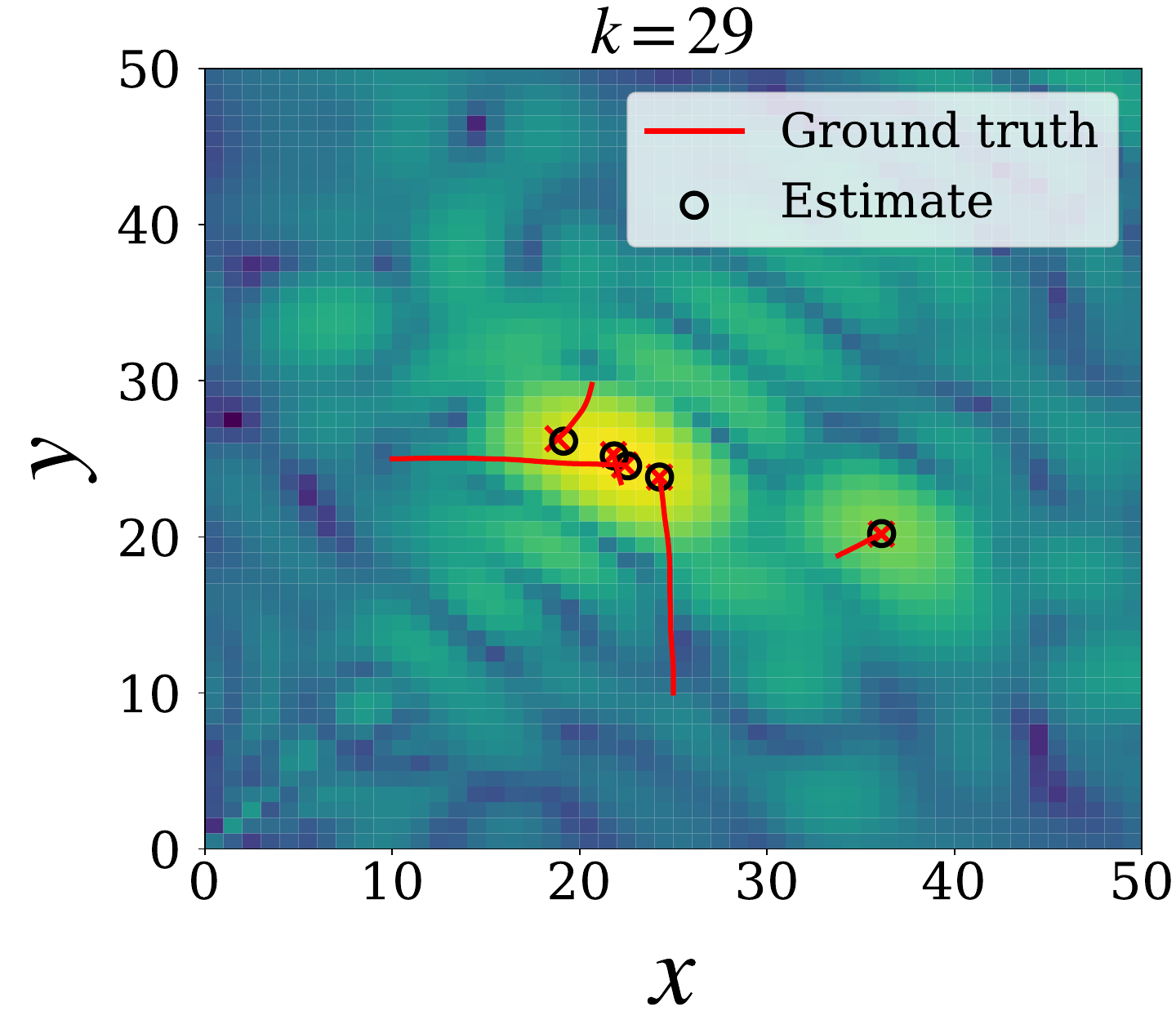}
	\end{subfigure}
	
	\vspace{0.8em}
	\begin{subfigure}[b]{0.49\linewidth}
		\centering
		\includegraphics[width=1.7in]{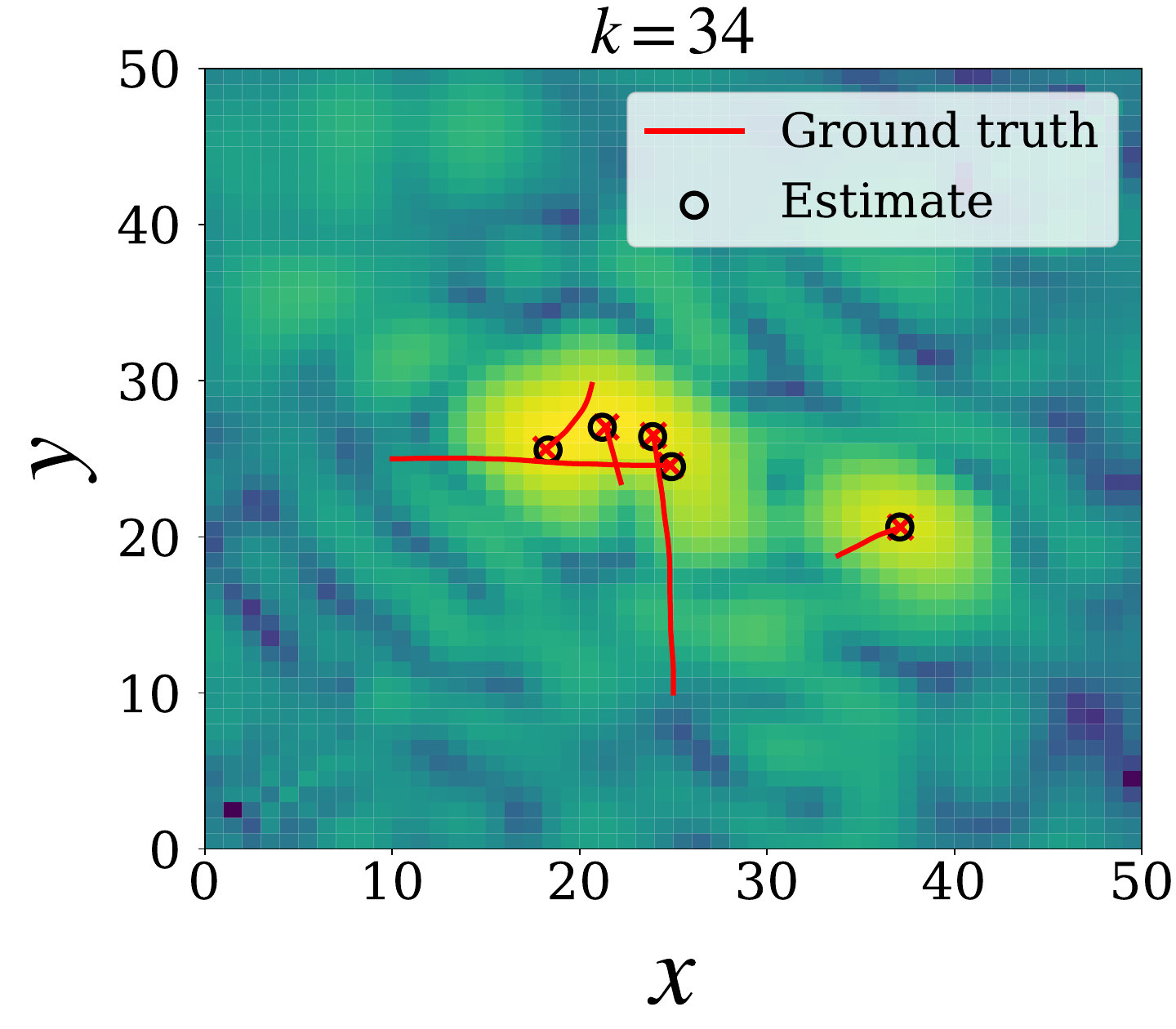}
	\end{subfigure}
	\hfill
	\begin{subfigure}[b]{0.49\linewidth}
		\centering
		\includegraphics[width=1.7in]{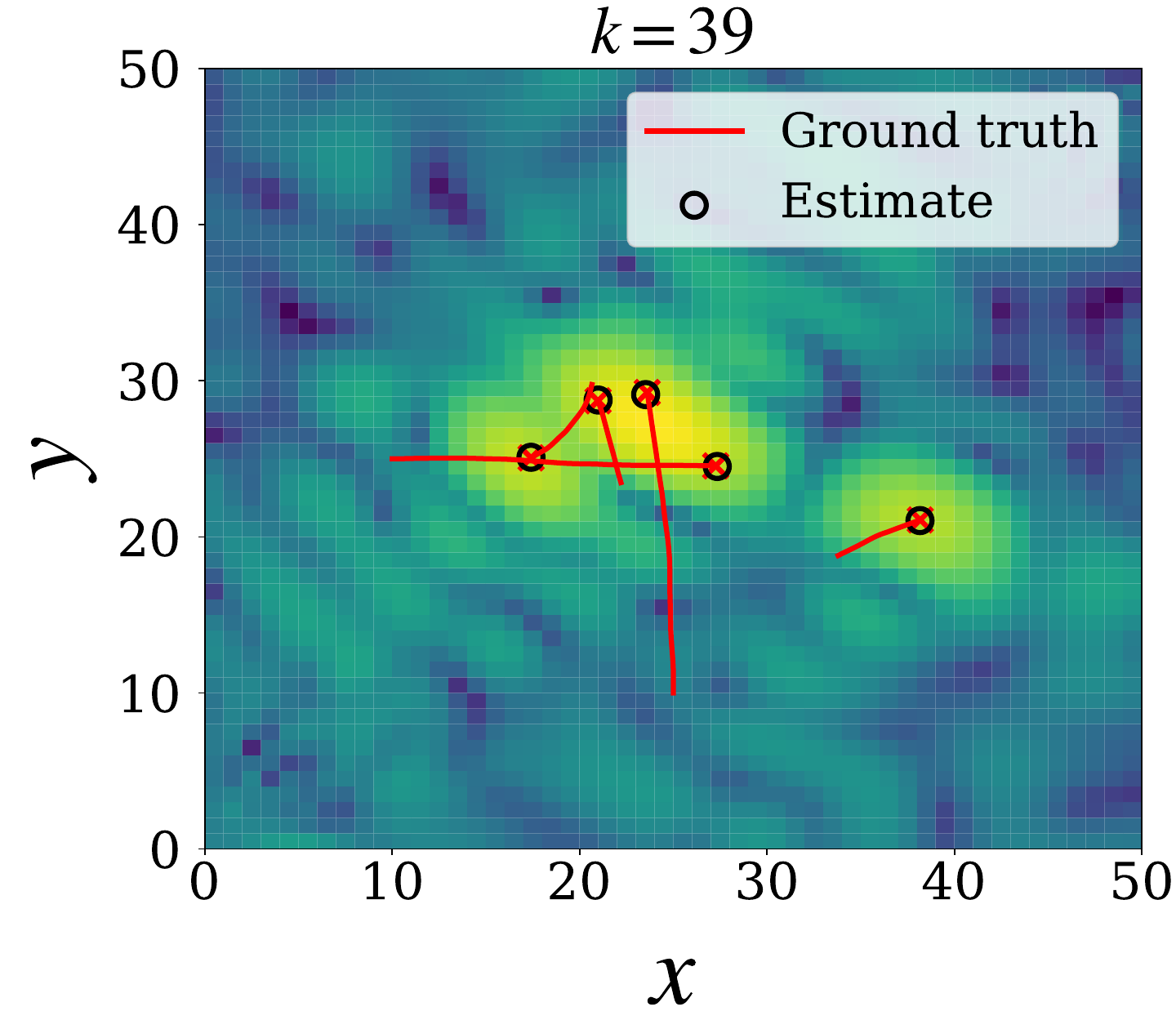}
	\end{subfigure}
	\caption{Representative tracking snapshots at time steps $k=25$, $30$, $35$, and $40$, shown from top left to bottom right. Red crosses and black circles denote the true and estimated object positions at the current time step, respectively. Red curves show the true historical trajectories. The background shows the Bartlett spectrum obtained by matched filtering.}
\label{track_vis}
	\vspace{-4mm}
\end{figure}

A visualization of the considered scenario is shown in Fig.~\ref{track_vis}. It can be seen that the synthetic radar has relatively low spatial resolution, making it challenging for conventional DTT to resolve closely spaced objects. In contrast, our proposed method can effectively distinguish and track nearby objects.

\begin{figure}[!t]
	\centering
	\includegraphics[width=3.5in]{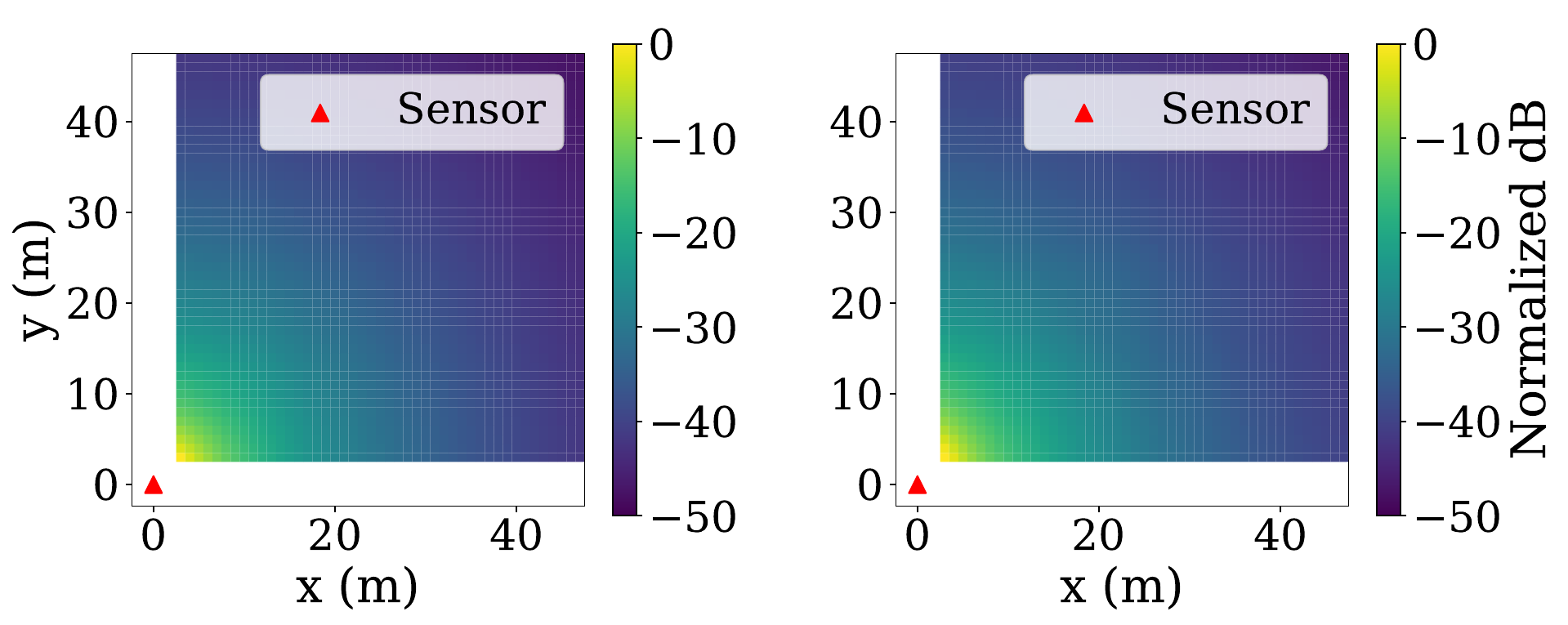}
	\caption{Radar intensity maps. Left: ground-truth intensity $\big|\varrho^{(i)}\big|^2$ computed from the radar equations used for data generation. Right: mean of the learned prior intensity PDF $p(\gamma;\mathbf{c})$ produced by the proposed neural-enhanced TkBD method\vspace{-2mm}.}
\label{I0}
\end{figure}

\begin{figure}[!t]
	\centering
	\includegraphics[scale=0.3]{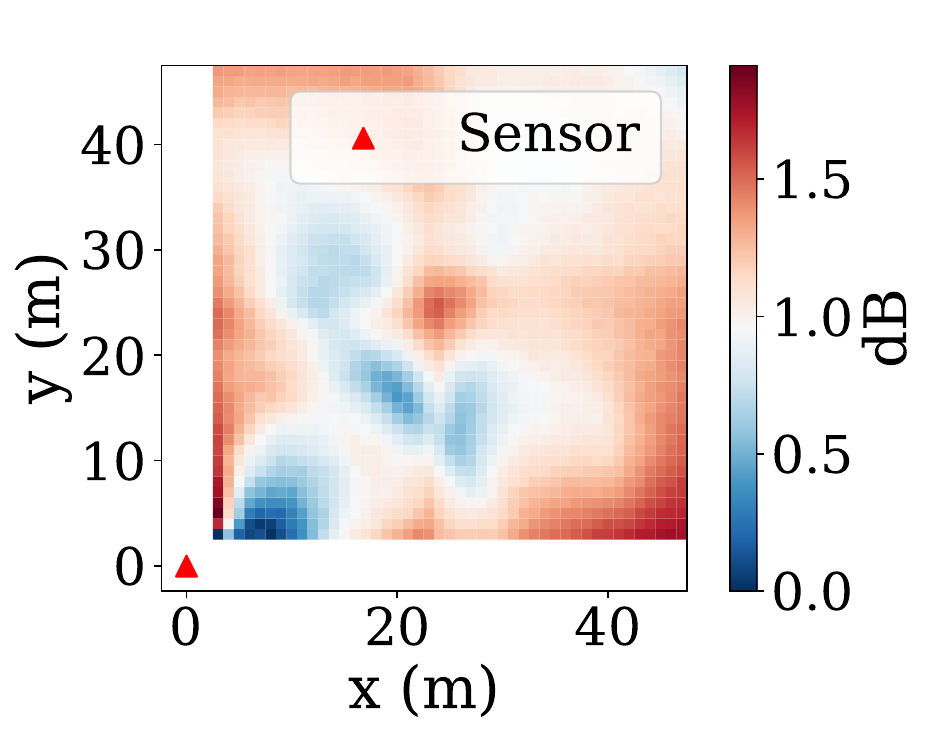}
\caption{Pointwise difference between the ground-truth radar-equation intensity and the learned prior intensity shown in Fig.~\ref{I0}. The small residuals indicate that the proposed neural-enhanced TkBD method accurately learns the radar-intensity model online\vspace{-2mm}.}
\label{I25}
	\vspace{-1.5mm}
\end{figure}
To visualize the prior PDF of the intensity online, we discretize the map into a uniform grid with spatial resolution $0.5 \times 0.5$\,m$^2$. For each grid cell, we draw 512 samples from the prior PDF $p(\gamma;\V{c})$, and report the sample mean. Figs.~\ref{I0} and \ref{I25} compare the resulting maps with the radar equation in \eqref{eq:radar_eq} used for data generation.  The small residuals in Fig.~\ref{I25} indicate that the proposed neural-enhanced TkBD method accurately learns the radar-intensity model during runtime\vspace{0mm}.

\section{Conclusion}

We proposed a neural-enhanced TkBD framework for MOT directly from complex-valued raw sensor measurements. The method augments a physics-based superpositional measurement model with an environment-dependent latent intensity variable that captures signal-power variations caused by unmodeled propagation effects. The prior PDF of this variable is represented by a normalizing flow and learned online in an unsupervised manner within the sequential Bayesian inference procedure. The learning objective is derived from an ELBO-based variational formulation, and the required expectations are evaluated using particle-based posterior representations produced by BP/SMC inference.

Numerical results in a synthetic multi-sensor radar tracking scenario demonstrate that the proposed method improves GOSPA and localization performance compared with DTT baselines and conventional TkBD. The results also show that the learned intensity model closely matches the radar-equation-based signal-power model used for data generation. This indicates that the proposed approach can exploit raw measurements not only for object tracking, but also for learning useful propagation-dependent structure online. Future work will focus on validation with real sensing data and extensions to more complex propagation environments.

\section{Acknowledgement}
This work was supported by the National Science Foundation (NSF) under CAREER Award No. 2146261.
\selectfont

\renewcommand{\baselinestretch}{0.99}
\IEEEtriggeratref{31}
\bibliographystyle{IEEEtran}
\bibliography{IEEEabrv, SALBooks, SALPapers,Temp}

\begin{thebibliography}{10}
\providecommand{\url}[1]{#1}
\csname url@samestyle\endcsname
\providecommand{\newblock}{\relax}
\providecommand{\bibinfo}[2]{#2}
\providecommand{\BIBentrySTDinterwordspacing}{\spaceskip=0pt\relax}
\providecommand{\BIBentryALTinterwordstretchfactor}{4}
\providecommand{\BIBentryALTinterwordspacing}{\spaceskip=\fontdimen2\font plus
\BIBentryALTinterwordstretchfactor\fontdimen3\font minus
  \fontdimen4\font\relax}
\providecommand{\BIBforeignlanguage}[2]{{%
\expandafter\ifx\csname l@#1\endcsname\relax
\typeout{** WARNING: IEEEtran.bst: No hyphenation pattern has been}%
\typeout{** loaded for the language `#1'. Using the pattern for}%
\typeout{** the default language instead.}%
\else
\language=\csname l@#1\endcsname
\fi
#2}}
\providecommand{\BIBdecl}{\relax}
\BIBdecl

\bibitem{BarWilTia:B11}
Y.~Bar-Shalom, P.~K. Willett, and X.~Tian, \emph{{Tracking and Data Fusion: A
  Handbook of Algorithms}}.\hskip 1em plus 0.5em minus 0.4em\relax Storrs, CT:
  Yaakov Bar-Shalom, 2011.

\bibitem{Mah:B07}
\vspace{0mm} R.~Mahler, \emph{{Statistical Multisource-Multitarget Information
  Fusion}}.\hskip 1em plus 0.5em minus 0.4em\relax Norwood, MA: Artech House,
  2007.

\bibitem{Bla:J04}
S.~S. Blackman, ``Multiple hypothesis tracking for multiple target tracking,''
  \emph{{IEEE} Trans. Aerosp. Electron. Syst.}, vol.~19, pp. 5--18, Jan. 2004.

\bibitem{Wil:J15}
J.~L. Williams, ``{Marginal multi-Bernoulli filters: RFS derivation of MHT,
  JIPDA and association-based MeMBer},'' \emph{{IEEE} Trans. Aerosp. Electron.
  Syst.}, vol.~51, no.~3, pp. 1664--1687, Jul. 2015.

\bibitem{MeyWil:J21}
F.~Meyer and J.~L. Williams, ``Scalable detection and tracking of geometric
  extended objects,'' \emph{{IEEE} Trans. Signal Process.}, vol.~69, pp.
  6283--6298, Oct. 2021.

\bibitem{Rei:J79}
D.~B. Reid, ``An algorithm for tracking multiple targets,'' \emph{{IEEE} Trans.
  Autom. Control}, vol.~24, no.~6, pp. 843--854, Dec. 1979.

\bibitem{BloBl:95}
E.~Bloem and H.~Blom, ``Joint probabilistic data association methods avoiding
  track coalescence,'' in \emph{Proc. CDC-95}, vol.~3, 1995, pp. 2752--2757
  vol.3.

\bibitem{HabThaThaMalKir:J13}
B.~Habtemariam, R.~Tharmarasa, T.~Thayaparan, M.~Mallick, and T.~Kirubarajan,
  ``{A multiple-detection joint probabilistic data association filter},''
  \emph{{IEEE} J. Sel. Topics Signal Process.}, vol.~7, no.~3, pp. 461--471,
  Jun. 2013.

\bibitem{LunGraOrg:J13}
C.~Lundquist, K.~Granstr\"om, and U.~Orguner, ``{An extended target CPHD filter
  and a gamma Gaussian inverse Wishart implementation},'' \emph{{IEEE} J. Sel.
  Topics Signal Process.}, vol.~7, no.~3, pp. 472--483, Jun. 2013.

\bibitem{ReuVoVoDie:J14}
S.~Reuter, B.-T. Vo, B.-N. Vo, and K.~Dietmayer, ``The labeled
  multi-{B}ernoulli filter,'' \emph{{IEEE} Trans. Signal Process.}, vol.~62,
  no.~12, pp. 3246--3260, 2014.

\bibitem{VoMa:J06}
B.~N. Vo and W.~K. Ma, ``The {G}aussian mixture probability hypothesis density
  filter,'' \emph{{IEEE} Trans. Signal Process.}, vol.~54, no.~11, pp.
  4091--4104, 2006.

\bibitem{GarWilGraSve:J18}
{\'A}.~F. Garc{\'\i}a-Fern{\'a}ndez, J.~L. Williams, K.~Granstr{\"o}m, and
  L.~Svensson, ``Poisson multi-{Bernoulli} mixture filter: Direct derivation
  and implementation,'' \emph{{IEEE} Trans. Aerosp. Electron. Syst.}, vol.~54,
  no.~4, pp. 1883--1901, Feb. 2018.

\bibitem{MeyBraWilHla:15}
F.~Meyer, P.~Braca, P.~Willett, and F.~Hlawatsch, ``{Scalable multitarget
  tracking using multiple sensors: A belief propagation approach},'' in
  \emph{Proc. {FUSION}-15}, 2015, pp. 1778--1785.

\bibitem{MeyKroWilLauHlaBraWin:J18}
F.~Meyer, T.~Kropfreiter, J.~L. Williams, R.~Lau, F.~Hlawatsch, P.~Braca, and
  M.~Z. Win, ``Message passing algorithms for scalable multitarget tracking,''
  \emph{Proc. {IEEE}}, vol. 106, no.~2, pp. 221--259, Feb. 2018.

\bibitem{LiaMey:J23}
M.~Liang and F.~Meyer, ``Neural enhanced belief propagation for multiobject
  tracking,'' \emph{{IEEE} Trans. Signal Process.}, vol.~72, pp. 15--30, Jan.
  2024.

\bibitem{JanMeySnyWigBauHil:J23}
J.~Jang, F.~Meyer, E.~R. Snyder, S.~M. Wiggins, S.~Baumann-Pickering, and J.~A.
  Hildebrand, ``Bayesian detection and tracking of odontocetes in {3-D} from
  their echolocation clicks,'' \emph{J. Acoust. Soc.}, vol. 153, no.~5, pp.
  2690--2705, 2023.

\bibitem{WeiLiaMey:J26}
S.~Wei, M.~Liang, and F.~Meyer, ``Bayesian multiobject tracking with
  neural-enhanced motion and measurement models,'' \emph{{IEEE} Trans. Signal
  Process.}, pp. 1--16, 2026.

\bibitem{Bar:J85}
Y.~Barniv, ``Dynamic programming solution for detecting dim moving targets,''
  \emph{{IEEE} Trans. Aerosp. Electron. Syst.}, vol.~21, no.~1, pp. 144--156,
  Jan. 1995.

\bibitem{MoySpaLam:J11}
L.~R. Moyer, J.~Spak, and P.~Lamanna, ``A multi-dimensional {Hough}
  transform-based track-before-detect technique for detecting weak targets in
  strong clutter backgrounds,'' \emph{{IEEE} Trans. Aerosp. Electron. Syst.},
  vol.~47, no.~4, pp. 3062--3068, Oct. 2011.

\bibitem{DelBelBer:J21}
C.~De~Lima, D.~Belot, R.~Berkvens, A.~Bourdoux, D.~Dardari, M.~Guillaud,
  M.~Isomursu, E.~S. Lohan, Y.~Miao, A.~Noll~Barreto \emph{et~al.},
  ``{Convergent communication, sensing and localization in 6G systems: An
  overview of technologies, opportunities and challenges},'' \emph{IEEE
  Access}, vol.~9, pp. 17\,180--17\,202, 2021.

\bibitem{LiuMasPet:J22}
F.~Liu, Y.~Cui, C.~Masouros, J.~Xu, T.~X. Han, Y.~C. Eldar, and S.~Buzzi,
  ``Integrated sensing and communications: Toward dual-functional wireless
  networks for {6G} and beyond,'' \emph{{IEEE} J. Sel. Areas Commun.}, vol.~40,
  no.~6, pp. 1728--1767, 2022.

\bibitem{KadBehSorWelAmiYerYoo:22}
S.~Kadambi, A.~Behboodi, J.~B. Soriaga, M.~Welling, R.~Amiri, S.~Yerramalli,
  and T.~Yoo, ``Neural {RF SLAM} for unsupervised positioning and mapping with
  channel state information,'' in \emph{Proc. IEEE ICC-22}, Aug. 2022, pp.
  3238--3244.

\bibitem{LiaLeiMey:J25}
M.~Liang, E.~Leitinger, and F.~Meyer, ``Direct multipath-based {SLAM},''
  \emph{{IEEE} Trans. Signal Process.}, vol.~73, pp. 2336--2352, Mar. 2025.

\bibitem{LiaMey:J26}
\BIBentryALTinterwordspacing
M.~Liang and F.~Meyer, ``Message passing for track-before-detect,'' 2026.
  [Online]. Available: \url{https://arxiv.org/abs/2401.12345}
\BIBentrySTDinterwordspacing

\bibitem{VoVo:J13}
B.-T. Vo and B.-N. Vo, ``Labeled random finite sets and multi-object conjugate
  priors,'' \emph{{IEEE} Trans. Signal Process.}, vol.~61, no.~13, pp.
  3460--3475, 2013.

\bibitem{DavGae:B18}
S.~J. Davey and H.~X. Gaetjens, \emph{{Track-Before-Detect Using Expectation
  Maximisation}}.\hskip 1em plus 0.5em minus 0.4em\relax Singapore: Springer,
  2018.

\bibitem{BuzLopVenFer:J08}
S.~Buzzi, M.~Lops, L.~Venturino, and M.~Ferri, ``{Track-before-detect
  procedures in a multi-target environment},'' \emph{{IEEE} Trans. Aerosp.
  Electron. Syst.}, vol.~44, no.~3, pp. 1135--1150, 2008.

\bibitem{ShlEld:J23}
N.~Shlezinger and Y.~C. Eldar, ``Model-based deep learning,'' \emph{Found.
  Trends Signal Process.}, vol.~17, no.~4, pp. 291--416, 08 2023.

\bibitem{KobPriBru:J21}
I.~Kobyzev, S.~J.~D. Prince, and M.~A. Brubaker, ``Normalizing {F}lows: An
  introduction and review of current methods,'' \emph{{IEEE} Trans. Pattern
  Anal. Mach. Intell.}, vol.~43, no.~11, pp. 3964--3979, 2021.

\bibitem{Kay:B93}
S.~M. Kay, \emph{{Fundamentals of Statistical Signal Processing: Estimation
  Theory}}.\hskip 1em plus 0.5em minus 0.4em\relax Prentice-Hall, 1993.

\bibitem{Bis:B06}
C.~M. Bishop, \emph{Pattern Recognition and Machine Learning}.\hskip 1em plus
  0.5em minus 0.4em\relax Berlin, Heidelberg: Springer-Verlag, 2006.

\bibitem{PapPil:02}
A.~Papoulis and S.~U. Pillai, \emph{Probability, Random Variables and
  Stochastic Processes}, 4th~ed.\hskip 1em plus 0.5em minus 0.4em\relax
  McGraw-Hill, 2002.

\bibitem{ShaKirLi:B02}
Y.~Bar-Shalom, T.~Kirubarajan, and X.-R. Li, \emph{{Estimation with
  Applications to Tracking and Navigation}}.\hskip 1em plus 0.5em minus
  0.4em\relax New York, NY: Wiley, 2002.

\bibitem{RahGarSve:17}
A.~S. Rahmathullah, {\'A}.~F. Garc{\'\i}a-Fern{\'a}ndez, and L.~Svensson,
  ``Generalized optimal sub-pattern assignment metric,'' in \emph{Proc.
  FUSION-17}, Jul. 2017, pp. 1--8.

\bibitem{MalZha:J93}
S.~Mallat and Z.~Zhang, ``Matching pursuits with time-frequency dictionaries,''
  \emph{{IEEE} Trans. Signal Process.}, vol.~41, no.~12, pp. 3397--3415, Dec.
  1993.

\bibitem{MeyBraWilHla:J17}
F.~Meyer, P.~Braca, P.~Willett, and F.~Hlawatsch, ``{A scalable algorithm for
  tracking an unknown number of targets using multiple sensors},'' \emph{{IEEE}
  Trans. Signal Process.}, vol.~65, no.~13, pp. 3478--3493, Jul. 2017.

\bibitem{WipRao:J07}
D.~P. {Wipf} and B.~D. {Rao}, ``{An empirical Bayesian strategy for solving the
  simultaneous sparse approximation problem},'' \emph{{IEEE} Trans. Signal
  Process.}, vol.~55, no.~7, pp. 3704--3716, Jul. 2007.

\bibitem{GerMecChrXenNan:J16}
P.~Gerstoft, C.~F. Mecklenbräuker, A.~Xenaki, and S.~Nannuru, ``Multisnapshot
  sparse bayesian learning for {DOA},'' \emph{{IEEE} Signal Process. Lett.},
  vol.~23, no.~10, pp. 1469--1473, Aug. 2016.

\end{thebibliography}

\end{document}